\documentclass{PoS}

\title{$H^\pm\to cb$  in models with two or more Higgs doublets}

\ShortTitle{$H^\pm\to cb$ in MHDMs}

\author{\speaker{S. Moretti}\thanks{SM thanks the Workshop organisers and the NExT Institute for financial support.}\\
        School of Physics and Astronomy, University of Southampton, \\
        Highfield, Southampton SO17 1BJ, United Kingdom\\
        E-mail: \email{S.Moretti@soton.ac.uk}}

\author{A.G. Akeroyd\\
        School of Physics and Astronomy, University of Southampton, \\
        Highfield, Southampton SO17 1BJ, United Kingdom\\
        E-mail: \email{A.G.Akeroyd@soton.ac.uk}}

\author{J. Hern\'andez-S\'anchez\\
       Facultad de Ciencias de la Electr\'onica, Benem\'erita Universidad Aut\'onoma de
       Puebla 
       and Dual C-P Institute of High Energy Physics, Apdo. Postal 542, 72570 Puebla, Puebla, M\'exico\\
       E-mail: \email{jaimeh@ece.buap.mx}}

\abstract{Searches for light $H^\pm$s via
$t\to H^\pm b$ are being carried out at the LHC. Herein, it is normally assumed that the dominant
decay channels are  $H^\pm\to \tau\nu$ and $H^\pm\to cs$ and separate data analyses 
are performed with comparable sensitivity to the underlying model assumptions. However, 
the $H^\pm\to cb$ decay rate can be as large as $80\%$  in models with two or more Higgs doublets 
with natural flavour conservation, while satisfying the constraint 
from $b\to s\gamma$ for $m_{H^\pm}< m_t$. Despite the current search strategy for
$H^\pm\to cs$ is also sensitive to $H^\pm\to cb$, a significant  gain in 
sensitivity could be obtained by tagging the $b$ quark from the decay $H^\pm\to cb$.}

\FullConference{Prospects for Charged Higgs Discovery at Colliders - CHARGED 2014,\\
		16-18 September 2014\\
		Uppsala University, Sweden}

\begin{document}

\section{Introduction}

\noindent
At the Large Hadron Collider (LHC),
if $m_{H^\pm} <  m_t$, $H^\pm$  states would mostly \cite{Aoki:2011wd} be produced in 
$t\to H^\pm b$ decays \cite{tbH}.
Searches in this channel are being performed by the LHC  experiments, assuming 
the decay modes $H^\pm\to cs$ and  $H^\pm\to \tau\nu$. 
Since no signal has been observed, constraints are obtained on the parameter space of 
a variety of models, chiefly 2-Higgs Doublet Models (2HDMs) \cite{Branco:2011iw}. 
Searches in these channels so far carried out at the LHC include:
1) $H^\pm\to cs$ with 4.7 fb$^{-1}$  by ATLAS \cite{ATLAS:search} and with 19.7 fb$^{-1}$ by CMS \cite{CMS:search};
2) $H^\pm\to \tau\nu$ with 19.5 fb$^{-1}$ by ATLAS \cite{ATLAS_Htau} and with 19.7 fb$^{-1}$ by CMS \cite{CMS_Htau}.
Although the current limits on $H^\pm\to cs$  can be applied to
the decay $H^\pm \to cb$ as well (as discussed in \cite{Logan:2010ag} in the Tevatron context), a further
improvement in sensitivity to  $t\to H^\pm b$ with $H^\pm \to cb$ could be obtained by tagging the 
$b$ quark which originates from $H^\pm$ \cite{Logan:2010ag,Akeroyd:1995cf,DiazCruz:2009ek}. 

We will estimate the increase in sensitivity to BR$(H^\pm \to cb)$  in a specific scenario, for definiteness, a 3-Higgs Doublet Model (3HDM) (see, e.g. \cite{Cree:2011uy})\footnote{As explained in \cite{Akeroyd:2012yg},  in the 
Aligned Two Higgs Doublet Model (A2HDM) \cite{Pich:2009sp} one can also have  a large BR$(H^\pm \to cb)$
 \cite{DiazCruz:2009ek} with $m_{H^\pm} < m_t$, so that our numerical results for the 3HDM apply directly to the A2HDM
too. 
In contrast, while large values of BR$(H^\pm \to cb)$ are also possible in the so called Type III 2HDM  \cite{Logan:2010ag,Akeroyd:1994ga,Aoki:2009ha}, they only occur for  $m_{H^\pm} > m_t$ due to the 
constraints from $b\to s \gamma$  requiring $m_{H^\pm} > 300$ GeV \cite{Hou:1987kf,Borzumati:1998tg,Misiak:2006zs}.
Finally, in the three other versions of the 2HDM (Type I, II and IV), in which 
 {BR($H^\pm \to \tau\nu$)} and {BR($H^\pm \to cs$)} dominate, one has that
 {BR($H^\pm\to cb)$} is always {$< 1\%$} (due to a small {$V_{cb}$}).}.
Reasons to consider a 3HDM could be the following: 
1) the existence already of 3 generations of quarks and leptons;
2) (scalar) dark matter (in presence of inert Higgs doublets) and a non-SM like sector.

\section{Charged Higgs bosons in the 3HDM}

\noindent
We will consider here the {`democratic' 3HDM} \cite{Cree:2011uy} wherein the fermionic states
  {$u,d,\ell$} obtain mass from {$v_u,v_d,v_\ell$} (the three different Vacuum Expectation Values (VEVs)), respectively.
 The mass matrix of the charged scalars
is diagonalised by the {$3\times 3$} matrix unitary {$U$}:
\begin{equation}
\left( \begin{array}{c} G^+ \\ H_2^+ \\ H_3^+ \end{array} \right) 
= U \left( \begin{array}{c} \phi_d^+ \\ \phi_u^+ \\ \phi_\ell^+ \end{array} \right).
\end{equation}
Henceforth, we will assume {$H^\pm_2$} to be the lightest state and relabel it as {$H^\pm$}.

The Yukawa couplings of the {$H^\pm$} in a 3HDM are given through the following Lagrangian
\begin{equation}
{\cal L}_{H^\pm} =
-\left\{\frac{\sqrt2V_{ud}}{v}\overline{u}
\left(m_d { X}{P}_R+m_u { Y}{P}_L\right)d\,H^+
+\frac{\sqrt2m_e }{v} { Z}\overline{\nu_L^{}}\ell_R^{}H^+
+{H.c.}\right\}.
\end{equation}
 In a 3HDM, {$X$},  {$Y$} and 
{$Z$} are defined in terms of the matrix elements of   {$U$},
\begin{equation}{
X=\frac{U_{12}}{U_{11}}, \;\,\;
Y=-\frac{U_{22}}{U_{21}}, \;\,\;
Z=\frac{U_{32}}{U_{31}}},
\end{equation}
\noindent
and
 are mildly constrained from the theoretical side, as the
 unitarity of  {$U$} leads to the relation 
\begin{equation}
 {|X|^2|U_{11}|^2+|Y|^2|U_{12}|^2+|Z|^2|U_{13}|^2=1}.
\end{equation}
Hence, the magnitudes of {$X$}, {$Y$} and 
{$Z$} cannot all be simultaneously {less or more than 1}.
 This is due to the fact that all three VEVs 
cannot be simultaneously large or small, as {$v^2_d+ v^2_u+ v^2_\ell=(246~{\rm GeV})^2$}.
Further theory constraints can be imposed via the usual requirements of $VV$ scattering unitarity ($V=W^\pm$ or $Z$), perturbativity, vacuum stability, positivity of mass eigenstates and of the Hessian, 
Electro-Weak Symmetry Breaking 
(EWSB) (now in presence of an $m_h=125~{\rm GeV}$ SM-like Higgs boson), etc. (see \cite{Keus:2014isa,Keus:2014jha,Keus:2013hya} for details), though all these
primarily affect the neutral Higgs sector of a 3HDM.

Indeed, are the 
phenomenological constraints those which impinge greatly on the allowed values of  {$X$, $Y$} and (less so) $Z$. The 
main limits come from the following low energy processes:
\begin{itemize}
\item {$Z\to b\overline b$:} 
{$|Y|< 0.72+0.24\left(\frac{m_{H^\pm}}{100 \rm GeV}\right)$}; 
\item {$b\to s\gamma$:} {$-1.1 < {\rm Re}(XY^*) < 0.7$}, e.g. for {$m_{H^\pm}=100$ GeV}.
\end{itemize}
\noindent
In essence, in
 the democratic 3HDM {{$H^\pm$} can be light} since {$XY^*$} is arbitrary.  As for LHC constraints enforced by the Higgs boson search
(and coupling measurements), these are rather loose as the $H^\pm$ state only enters via loop effects (e.g. in $\gamma\gamma$ and $Z\gamma$ decays).

\section{Results}

\noindent
In the light of the previous discussion, a
distinctive signal of the  $H^\pm$ boson from a 3HDM would then be
a large BR($H^\pm\to cb)$ with the charged Higgs boson emerging from an (anti)top decay (since $m_{H^\pm}<m_t$).
The necessary condition for this is: {$|X|>> |Y|,|Z|$}. (In the numerical analysis we fix {$m_{H^\pm}=120$ GeV} and 
{$|Z|=0.1$}.) We illustrate in Fig.~\ref{fig:BRs} the BR($H^\pm \to cb$) and
BR($H^\pm \to cs$) in a 3HDM. Over the strip between  the lines $|XY^*|=0.7$ and 1.1
(notice that this area does not correspond to the entire region surviving $b\to s\gamma$ constraints), it is
clear the predominance of the former over the latter. 
 
As mentioned, both ATLAS and CMS have searched for {$t\to H^\pm b$} and 
{$H^\pm\to cs$}. The procedure is simple. 
 Top quarks are produced in pairs  via {$q\bar q,gg\to t\overline t$}.
 One (anti)top then decays via {$t/\overline t\to W b$}, with 
$W\to e\nu$ or $\mu\nu$.
 The other  (anti)top decays via {$t/\overline t\to H^\pm b$}. Hence,
 {$H^\pm\to cs$} gives two (non-{$b$} quark) jets.
 Candidate signal events are therefore {$b\overline b e\nu$} plus two non-{$b$} 
jets.  A peak at $m_{H^{\pm}}$ in the invariant mass distribution of non-{$b$} jets is the hallmark signal. The 
 main background comes from  $t/\overline t\to W b$ and 
{$W\to ud/cs$}, which would give a peak at {$m_{W^\pm}$.

We simply remark here that 
applying a third {$b$}-tag would improve sensitivity to $H^\pm \to cb$ greatly, as  the main background from 
{$W\to cb$} has a very small rate. This is made explicit by choosing a
  {$b$}-tagging efficiency {$\epsilon_b=0.5$},
a {$c$}-quark mistagging rate  {$\epsilon_c=0.1$} and a
 light quark $(u,d,s)$ mistagging rate 
{$\epsilon_j=0.01$}. It follows  that the
 estimate gain in sensitivity is then:
\begin{equation}
\frac{[S/\sqrt B]_{\rm btag}}{[S/\sqrt B]_{\rm \not{btag}}}
\sim \frac{\epsilon_b\sqrt 2}{\sqrt{(\epsilon_j+\epsilon_c)}}\sim {2.13}.
\end{equation}

\begin{figure}
\begin{center}
\includegraphics[width=5.5 cm, angle=0]{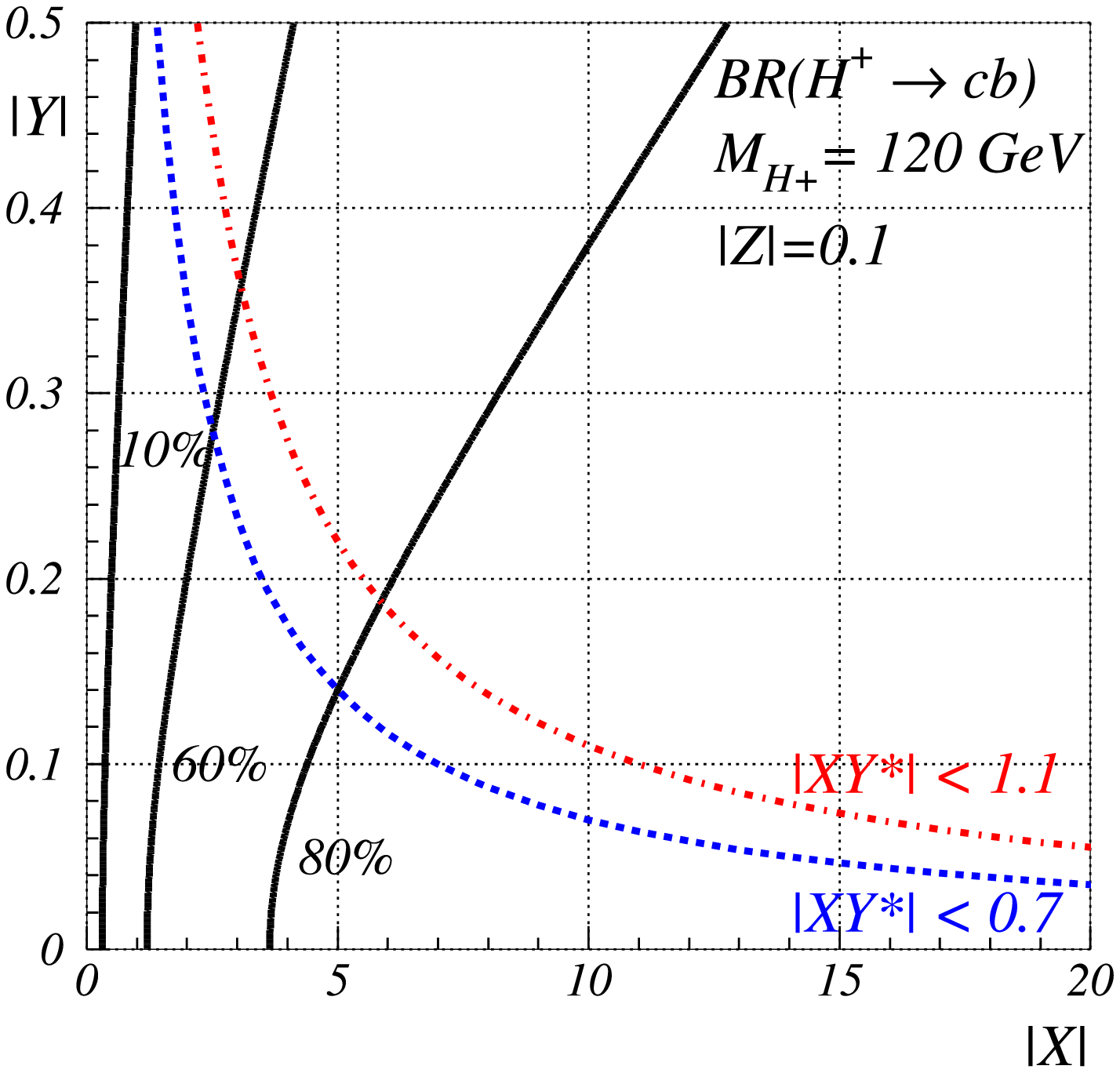}
\includegraphics[width=5.5 cm, angle=0]{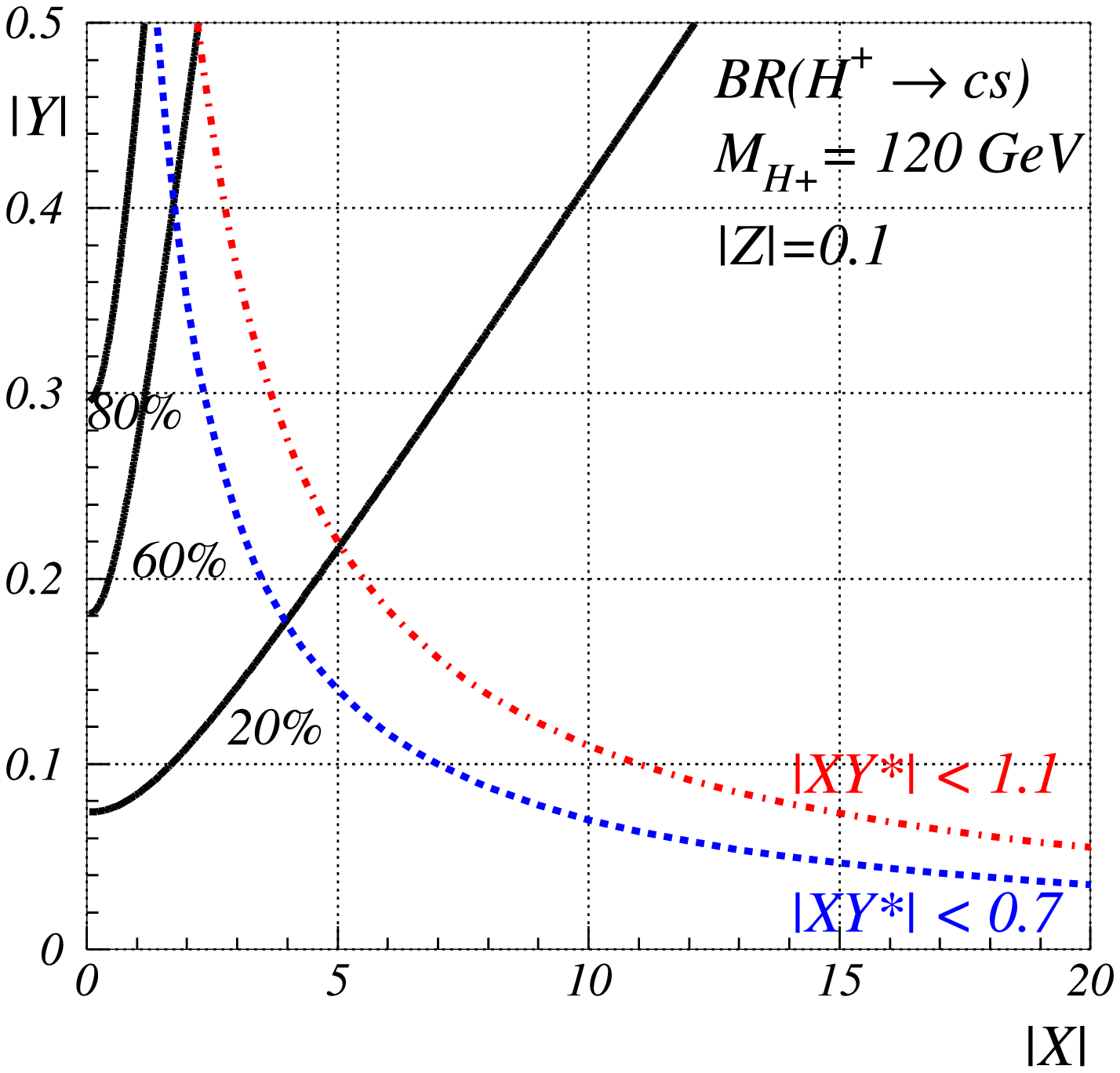}
\caption{Left: BR($H^\pm \to cb$) in the plane $[|X|,|Y|]$. Right: BR($H^\pm \to cs$) over the same plane.}
\label{fig:BRs}
\end{center}
\end{figure}

Current ATLAS  and CMS limits for {$m_{H^\pm}=120$ GeV} are of order
{BR$(t\to H^\pm b) < 0.02$} (assuming {BR$(H^\pm\to cs)=100\%$}). 
 In the plane of {$[|X|,|Y|]$} we now show contours of:
1) {BR($t\to H^\pm b$)$\times$BR($H^\pm\to cb$ + $cs$});
2) {BR$(t\to H^\pm b$)$\times$BR$(H^\pm\to cb)$}. This is done in Fig.~\ref{fig:contours}, 
from where it is clearly visible that
 constraints from {$t\to H^\pm b$}
are competitive with those from {$b\to s\gamma$}. In fact,
 {BR($t\to H^\pm b)<2\%$} rules out two regions which cannot be excluded via 
{$b\to s\gamma$}:
1) {{$15 < |X| < 40 $ and $0< |Y|<0.04$}};
2) {{$0< |X|<4$ and $0.3 > |Y|> 0.8$}}. Further,
 tagging the {$b$}-quark from {$H^\pm\to cb$}  would possibly allow sensitivity to
{BR($t\to H^\pm b)<0.5\%$} or less so that
 {$t\to H^\pm b$} combined with {$H^\pm\to cb$} could
provide even stronger constraints on the {$[|X|,|Y|]$} plane
(or perhaps enable discovering {$H^\pm \to cb$}).

\begin{figure}
\begin{center}
\includegraphics[width=5.5 cm, angle=0]{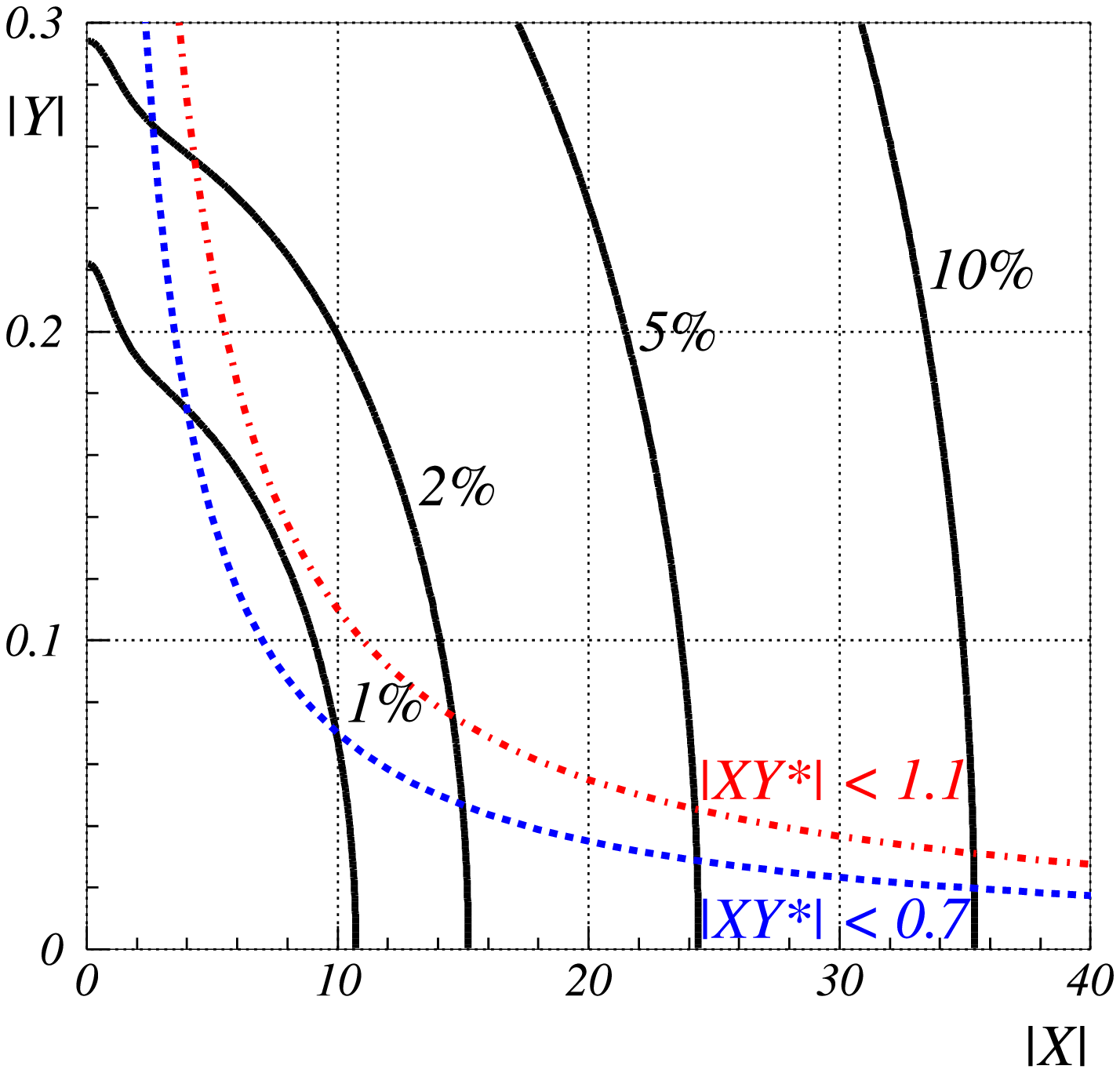}
\includegraphics[width=5.5 cm, angle=0]{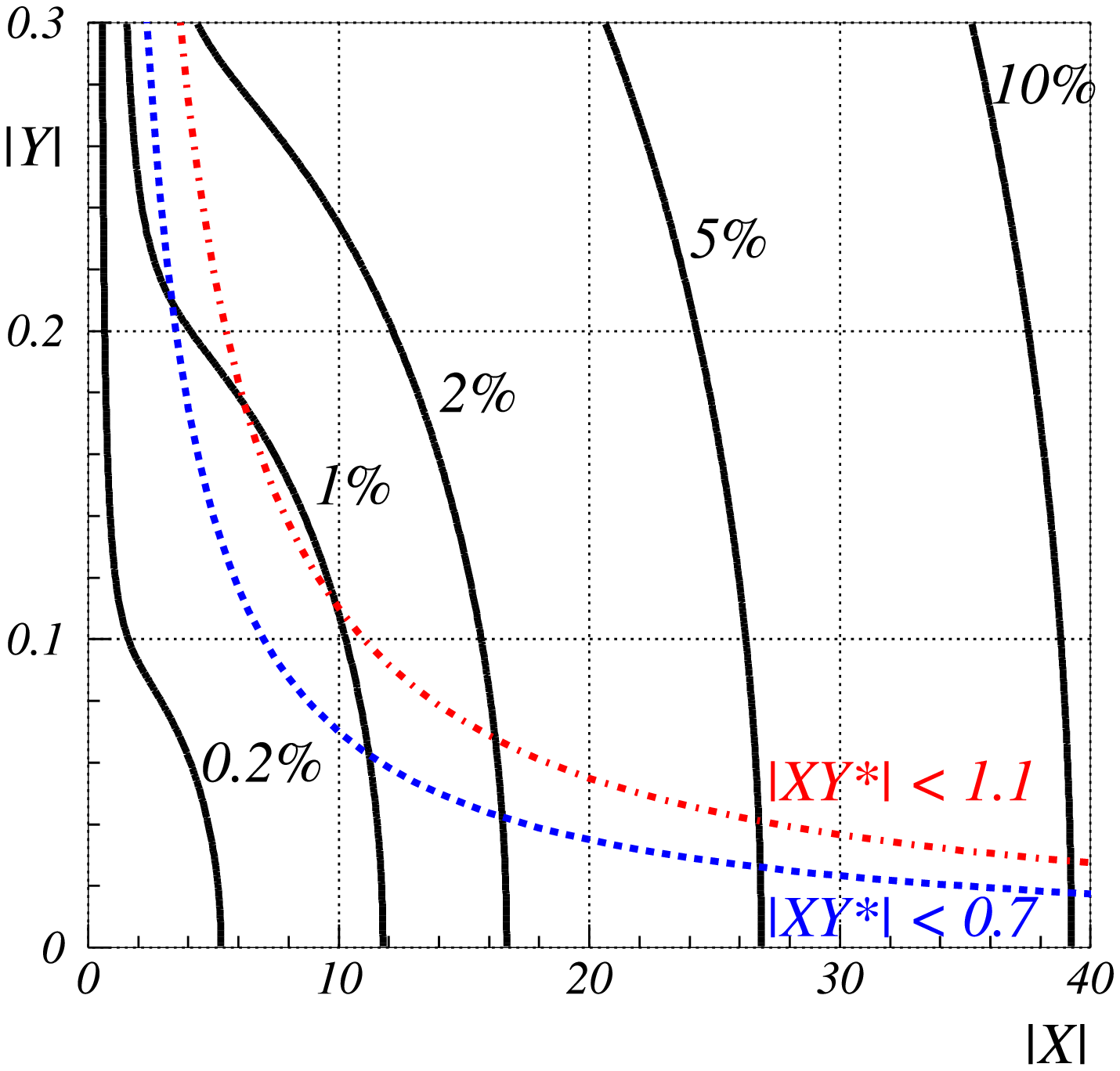}
\caption{{Left}: BR($t\to H^\pm b$)$\times$BR($H^\pm\to cb$ +
$cs$) (no $b$-tag). {Right}: BR($t\to H^\pm b$)$\times$BR($H^\pm\to cb$) ($b$-tag).}
\label{fig:contours}
\end{center}
\end{figure}

\section{Conclusions}

\noindent
A Higgs particle has been discovered, maybe there are more such states to be found, including a  $H^\pm$. 
We have emphasied here that a light (with mass below $m_t$) $H^\pm$ is possible in a 3HDM wherein  
  {$H^\pm\to cb$} can be dominant. Based on ongoing analyses by ATLAS and CMS searching for 
{$t\to H^\pm b$, $H^\pm\to cs$}, which are already sensitive
to   {$H^\pm\to cb$}, we proposed 
 {tagging the {$b$}-quark} from {$H^\pm\to cb$}, procedure that could further improve
sensitivity to the
fermionic couplings of  {$H^{\pm}$} ({$X$} and {$Y$}). This is a
 straightforward extension of ongoing searches for
{$t\to H^\pm b$ and $H^\pm\to cs$} that would enable one to make rather definitive statements regarding
the viability of  a 3HDM (and also a A2HDM).

\end{document}